# MEG-MASC: a high-quality magneto-encephalography dataset for evaluating natural speech processing.

**Laura Gwilliams**[1, 2, 3, *], **Graham Flick**[2, 3, 4], **Alec Marantz**[2, 3, 4], **Liina Pylkkänen**[2, 3, 4], **David Poeppel**[2, 5], **and Jean-Rémi King**[2, 6]

[1]Department of Neurological Surgery, University of California, San Francisco
[2]Department of Psychology, New York University, New York
[3]NYU Abu Dhabi Institute, Abu Dhabi
[4]Department of Linguistics, New York University, New York
[5]Ernst Struengmann Institute for Neuroscience, Frankfurt
[6]PSL University, Ecole Normale Supérieure, Paris
[*]leg5@nyu.edu

## ABSTRACT

The "MEG-MASC" dataset provides a curated set of raw magnetoencephalography (MEG) recordings of 27 English speakers who listened to two hours of naturalistic stories. Each participant performed two identical sessions, involving listening to four fictional stories from the Manually Annotated Sub-Corpus (MASC) intermixed with random word lists and comprehension questions. We time-stamp the onset and offset of each word and phoneme in the metadata of the recording, and organize the dataset according to the 'Brain Imaging Data Structure' (BIDS). This data collection provides a suitable benchmark to large-scale encoding and decoding analyses of temporally-resolved brain responses to speech. We provide the Python code to replicate several validations analyses of the MEG evoked related fields such as the temporal decoding of phonetic features and word frequency. All code and MEG, audio and text data are publicly available to keep with best practices in transparent and reproducible research.

## Background & Summary

Humans have the unique ability to produce and comprehend an infinite number of novel utterances. This capacity of the human brain has been the subject of vigorous studies for decades. Yet, the core computational mechanisms upholding this feat remain largely unknown [1, 2, 3].

To tackle this issue, a common experimental approach has been to decompose language processing into elementary computations using highly controlled factorial designs. This approach allows experimenters to compare average brain responses to carefully chosen stimuli and make inferences based on the select ways that those stimuli were designed to differ. The field has learnt a lot about the neurobiology of language by taking this approach; however, factorial designs also face several key challenges [4]. First, this method has lead the community to study language processing in atypical scenarios (*e.g.* using unusual text fonts [5], meaningless syntactic constructs [6, 7], or words and phrases isolated from context [8, 9]). Presenting language in this unconventional manner runs the risk of studying phenomena that are not representative of how language is naturally processed. Second, high-level cognitive functions can be difficult to fully orthogonalize in a factorial design. For instance, comparing brain responses to words and sentences matched in length, syntactic structure, plausibility and pronunciation is often close to impossible. In the best case, experimenters will be forced to make concessions on how well the critical contrasts are controlled. In the worst case, unidentified confounds may drive differences associated with experimental contrasts, leading to incorrect conclusions.

During the past decade, several studies have complemented the factorial paradigm with more natural environments. In these studies, participants listen to continuous speech [10, 11, 12], read continuous prose [13, 14] or watch videos that include verbal communication [15]. This approach is more likely to recruit neural computations that are representative of day-to-day language processing. Complications arising from correlated language features can be overcome by explicitly modelling properties of interest, in tandem with potential confounds. This allows variance belonging to either source to be appropriately distinguished.

To analyze the brain responses to the complex stimulation that natural language provides, a variety of encoding and decoding methods have proved remarkably effective [10, 16, 17, 18, 19, 20, 21]. Consequently, language studies based on naturalistic designs have since flourished [11, 22]. The popularity of this approach has some of its roots in the rise of natural language processing (NLP) algorithms, which map remarkably onto brain responses to written and spoken sentences [23, 24, 25, 26, 27, 28, 29, 30]. Such tools also allow experimenters to annotate the language stimuli for features of interest, without relying on time-consuming annotations done by hand. These data have allowed researchers to identify the main semantic components [10], recover the hierarchy of integration constants in the language network [31], distinguish syntax and semantics hubs [32] and to track the hierarchy of predictions elicited during speech processing [28, 33, 34]. More generally, brain responses to natural stories have proved useful in keeping participants engaged, while studying the neural representations of phonemes, word surprisal and entropy [22, 35, 36].

While large and high-quality functional Magnetic Resonance Imaging (fMRI) datasets related to language processing have recently been released [37, 38], there is currently little publicly available high-quality temporally-resolved brain recordings acquired during story listening. In particular, there is, to date, no public magneto-encephalography (MEG) with (1) several hours of story listening (2) multiple sessions (3) a systematic audio, phonetic and word annotations (4) a standardize data structure.

In the present study, 27 English-speaking subjects performed ≈two hours of story listening, punctuated by random word lists and comprehension questions in the MEG scanner. Except if stated otherwise, each subject listened to four distinct fictional stories twice.

## Methods

### Participants

Twenty-seven English-speaking adults were recruited from the subject pool of NYU Abu Dhabi (15 females; age: M=24.8, SD=6.4). All participants provided a written informed consent and were compensated for their time. Participants reported having normal hearing and no history of neurological disorders. All but one participants (S20) were native English speakers. All but five participants (S3, S11, S15, S19, S20) performed two identical one-hour-long sessions. These two recording sessions were separated by at least one day and at most two months depending on the availability of the experimenters and of the participants. The study was approved by the Institution Review Board (IRB) ethics committee of New York University Abu Dhabi.

### Procedure

Within each ∼1 h recording session, participants were recorded with a 208 axial-gradiometer MEG scanner built by the Kanazawa Institute of Technology (KIT), and sampled at 1,000 Hz, and online band-pass filtered between 0.01 and 200 Hz while they listened to four distinct stories through binaural tube earphones (Aero Technologies), at a mean level of 70 dB sound pressure level.

Before the experiment, participants were exposed to 20 sec of each of the distinct speaker voices used in the study to (i) clarify the structure of the session and (ii) familiarize the participants with these voices.

The order in which the four stories were presented was assigned pseudo-randomly, thanks to a "Latin-square design" across participants. This participant-specific order was used for both recording sessions.

To ensure that the participants were attentive to the stories, they answered, every ∼3 min and with a button press, a two-alternative forced-choice question relative to the story content (*e.g.* 'What precious material had Chuck found? Diamonds or Gold'). Participants performed this task with an average accuracy of 98%, confirming their engagement with and comprehension of the stories.

Participants who did not already have a T1-weighted anatomical scan usable for the present study were scanned in a 3T Magnetic-Resonance-Imaging (MRI) scanner after the MEG recording to avoid magnetic artefacts. Six participants did not return for their T1 scan.

Before each MEG session, the head shape of each participant was digitized with a hand-held FastSCAN laser scanner (Polhemus), and co-registered with five head-position coils. The positions of these coils with regard to the MEG sensors were collected before and after each recording and stored in the 'marker' file, following the KIT's system. The experimenter continuously monitored head position during the acquisition to ensure that the participants did not move.

### Stimuli

Four English fictional stories were selected from the Open American National Corpus [**?**] – a manually annotated corpus distributed without license or other restrictions[1]:

- 'Cable spool boy': a 1,948-word story narrating two young brothers playing in the woods

---
[1] https://anc.org/data/masc/corpus/577-2/

- 'LW1': a 861-word story narrating an alien spaceship trying to find its way home
- 'Black willow': a 4,652-word story narrating the difficulties an author encounters during writing.
- 'Easy money': a 3,541-word fiction narrating two friends using a magical trick to make money.

An audio track corresponding to each of these stories was synthesized using Mac OS Mojave's (c) text-to-speech. To help decorrelate language from acoustic representations, we varied both voices and speech rate every 5-20 sentences. Specifically, we used three distinct synthetic voices, namely 'Ava', 'Samantha' and 'Allison' speaking between 145 and 205 words per minute. Additionally, we varied the silence between sentences between 0 and 1,000 ms.

Each story was divided into ≈ 5,min sound files. In between these sounds – approximately every 30 s – we played a random word list generated from the unique content words (nouns, proper nouns, verbs, adverbs and adjectives) of the preceding 5 min segment presented in random order. In addition, a very small fraction (<1%) of non-words were introduced in natural sentences.

Hereafter, and following the BIDS labeling [39], each "task" corresponds to the concatenation of these sentences and word lists. Each subject listened to the exact same set of four tasks, only in a different block order.

### Preprocessing

**MEG**   The MEG dataset and its annotations are shared raw (i.e. not preprocessed) organized according to the Brain Imaging Data Structure [39] MNE-BIDS [40].

**MRI**   To avoid subject identification, the T1-weighted MRI anatomical scan was defaced using Freesurfer [41] [2] and manually checked.

The alignment between the spaces of (1) the head-position coils, (2) the MEG sensors and (3) the T1 MRI was co-registered manually with MNE-Python [42].

**Stimuli**   We include in the dataset: the original stories ('stimuli/text'), the stories intertwined with the word lists ('stimuli/text_with_wordlists'), their corresponding audio tracks ('stimuli/audio') and the alignment between the audio and the words and phonemes ('stimuli/events').

Both sentences and word lists were annotated for phoneme boundaries and labels (107 phonetic features) using the 'Gentle aligner' from the Python module lowerquality *https://lowerquality.com/gentle/*. However, the inclusion of the original audio leaves the possibility for future research to develop more advanced alignment technique and recover additional features.

For each phoneme and word, we indicate the corresponding voice, speech rate, wav file, story, word position within the sequence, and sequence position within the story, and whether the sequence is a word list or a sentence.

## Computing environment

In addition to the packages mentioned in this manuscript, the processing of the present data is based on the free and open-source ecosystem of the neuroimaging community. In particular, the, we used:

- MNE BIDS [40] (https://mne.tools/mne-bids)
- Bids-Validator  (https://github.com/bids-standard/bids-validator)
- Nibabel [43] (https://nipy.org/nibabel/)
- Scikit-Learn  [44](https://scikit-learn.org/)
- Pandas [45] (https://pandas.pydata.org/)

## Data Records

The dataset is organized according to Brain Imaging Data Structure (BIDS) 1.2.1 [39] and publicly available on the Open Science Framework data repository [3] under a Creative Common Licence 0. An image of the folder structure is provided in Figure 1. The detailed description of the BIDS file system is available at http://bids.neuroimaging.io/. In summary,

- './dataset_description.json' describes the dataset

---

[2] https://surfer.nmr.mgh.harvard.edu/fswiki/mri_deface
[3] https://osf.io/ag3kj/

- './participants.tsv' indicates the age and gender of each participant
- './stimuli/' contains the original texts, the modified texts (*i.e.* with word lists), the synthesized audio tracks and the alignment between text and audio.
- each './sub-SXXX' contains the brain recordings of a unique participant divided by session (*e.g.* 'ses-0' and 'ses-1')
- in each session folder lies the anatomical and the meg data.
- Sessions are numbered by temporal order.
- Tasks are numbered by a unique integer common across participants.

The dataset can be read directly with MNE-BIDS [40] 1.

## Technical Validation

The present dataset was validated by Bids-Validator [4].

MEG recordings are notoriously noisy and thus challenging to validate empirically. In particular, MEG can be corrupted by environmental noise (nearby electronic systems) and physiological noise (eye movement, heart activity, facial movements) [46]. To address this issue, several labs have proposed a myriad of preprocessing techniques based on temporal and spatial filtering [47] and trial and channel rejection [48]. However, there is currently no accepted standard for the selection and ordering these preprocessing steps. Consequently, we here opted for (1) a minimalist preprocessing pipeline derived from MNE-Python's default pipeline [42] followed by (2) average evoked responses and (3) standard single-trial linear decoding analyses.

### Minimal preprocessing

For each subject separately, and using the default parameters of MNE-Python, we:

- bandpass filtered the MEG data between 0.5 and 30.0 Hz,
- temporally-decimate the data 10x,
- segment these continuous signals between -200 ms and 600 ms after word and phoneme onset,
- apply a baseline correction between -200 ms
- and clip the MEG data between fifth and ninety-fifth percentile of the data across channels.

### Evoked

Figure 2 displays the median evoked related fields (ERFs) across participants and words onset and after phoneme onsets, respectively. Both of these topographies are typical of auditory activity in MEG [36].

### Decoding

For each recording independently, our objective was to verify the alignment between the word annotations and the MEG recordings. To this end, we trained a linear classifier $W \in R^d$ across all $d = 208$ magnetometers ($X \in R^{n \times d}$), for each time sample relative to word (or phoneme) onset independently, and for each subject separately. The classifier consisted of a standard scaler, followed by a linear ridge regression implemented by scikit-learn [44].

- decode the zipf-frequency of each word in the English language, as defined by the WordFreq package [49].
- decode whether the phoneme is voiced or not.

The decoding pipeline was trained and evaluated using a five-split cross-validation scheme (with shuffling). The scoring metric reported is Pearson R correlation.

The results displayed in Figure 3 show a reliable decoding at the phoneme and at the word level, across both subjects and tasks (*i.e.* stories).

---
[4]https://github.com/bids-standard/bids-validator

## Usage Notes

```
import pandas as pd
import mne_bids

bids_path = mne_bids.BIDSPath(
    subject='01',
    session='0',
    task='0',
    datatype="meg",
    root='my/data/path/',
)
raw = mne_bids.read_raw_bids(bids_path)
raw.load_data().data # channels X times

df = raw.annotations.to_data_frame()
```

Accessing all sound, word and phoneme annotation is directly readable in a Pandas [50] DataFrame format 4:

```
df = pd.DataFrame(df.description.apply(eval).to_list())
```

## Code availability

The code is available on https://github.com/kingjr/masc-bids.

## Acknowledgements

## Author contributions statement

L.G. and J-R.K. conceived the experiment. L.G. conducted the experiment. L.G. and J-R.K. analyzed the results. A.M., D.P. and J-R.K. contributed to the funding of the study. All authors reviewed the manuscript.

## Competing interests

(mandatory statement)

The corresponding author is responsible for providing a competing interests statement on behalf of all authors of the paper. This statement must be included in the submitted article file.

## Figures & Tables

```
├── README
├── dataset_description.json
├── participants.json
├── participants.tsv
├── stimuli
│   ├── audio
│   │   ├── cable_spool_fort_0.wav
│   │   ├── cable_spool_fort_1.wav
│   │   ├── ...
│   │   └── the_black_willow_9.wav
│   ├── text
│   │   ├── cable_spool_fort.txt
│   │   ├── easy_money.txt
│   │   ├── lw1.txt
│   │   └── the_black_willow.txt
│   └── text_with_wordlists
│       ├── cable_spool_fort_produced_0.txt
│       ├── cable_spool_fort_produced_1.txt
│       ├── ...
│       └── the_black_willow_produced_9.txt
├── sub-01
│   ├── ses-0
│   │   ├── anat
│   │   ├── meg
│   │   │   ├── sub-01_ses-0_acq-ELP_headshape.pos
│   │   │   ├── sub-01_ses-0_acq-HSP_headshape.pos
│   │   │   ├── sub-01_ses-0_coordsystem.json
│   │   │   ├── sub-01_ses-0_task-0_channels.tsv
│   │   │   ├── sub-01_ses-0_task-0_events.tsv
│   │   │   ├── sub-01_ses-0_task-0_markers.mrk
│   │   │   ├── sub-01_ses-0_task-0_meg.con
│   │   │   ├── sub-01_ses-0_task-0_meg.json
│   │   │   ├── sub-01_ses-0_task-1_channels.tsv
│   │   │   ├── sub-01_ses-0_task-1_events.tsv
│   │   │   ├── sub-01_ses-0_task-1_markers.mrk
│   │   │   ├── sub-01_ses-0_task-1_meg.con
│   │   │   ├── sub-01_ses-0_task-1_meg.json
│   │   │   ├── ...
│   │   └── sub-01_ses-0_scans.tsv
│   └── ses-1
│       ├── anat
│       ├── meg
│       │   ├── sub-01_ses-1_acq-ELP_headshape.pos
│       │   ├── ...
│       └── sub-01_ses-1_scans.tsv
├── sub-02
│   ├── ses-0
│   │   ├── anat
│   │   │   ├── sub-02_ses-0_FLASH.json
│   │   │   └── sub-02_ses-0_FLASH.nii.gz
│   │   ├── meg
│   │   │   ├── sub-02_ses-0_acq-ELP_headshape.pos
...
```

**Figure 1.** Dataset file structure

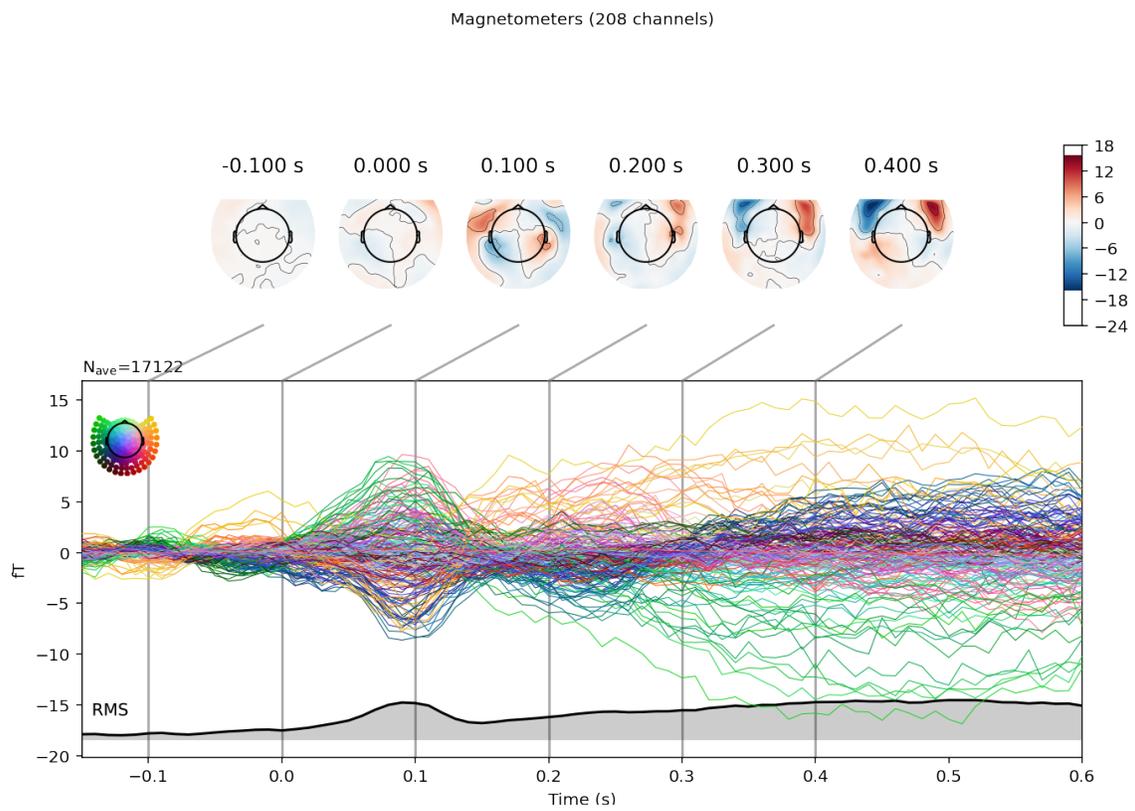

**Figure 2.** Average (across subjects) evoked response to all words (median). The gray area indicates the global field power (GFP).

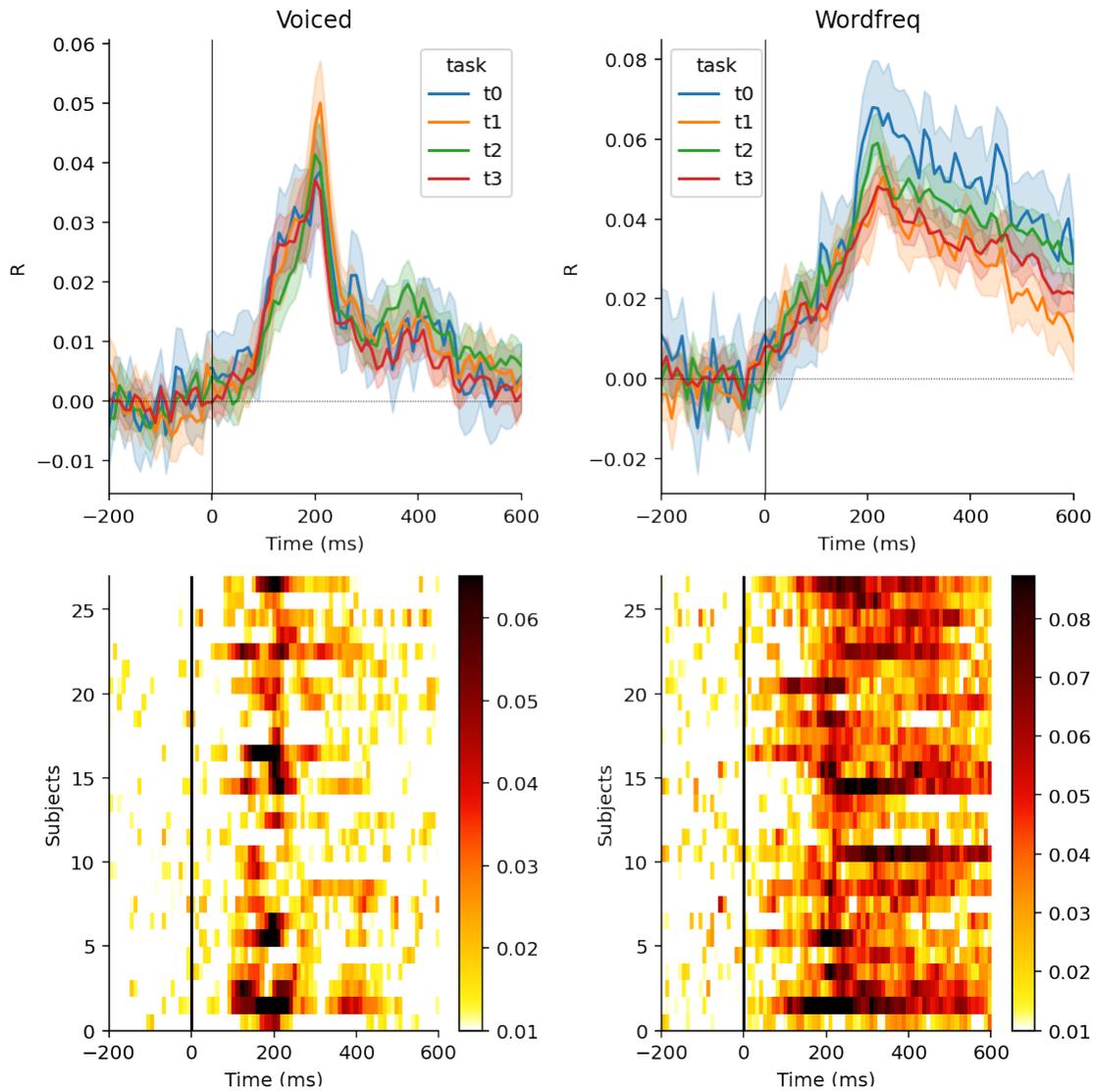

**Figure 3.** A. Average decoding of whether the phoneme is voiced or not as a function of time following phoneme onset. The four colors refer to the four tasks (stories+word lists). Error bar are SEM across subjects. B. Same as A for the decoding of words' zipf frequency as a function of word onset. C. Decoding of voicing (average across all tasks) for each participant, as a function of time following phoneme onset. D. Same as C for decoding of word frequency (average across all tasks) for each participant, as a function of time following word onset.

|  | start | kind | sound | phoneme | word | story | condition | speech_rate | voice |
|---|---|---|---|---|---|---|---|---|---|
| 0 | 0.00 | sound | stimuli/audio/lw1_0.0.wav | NaN | NaN | lw1 | NaN | NaN | NaN |
| 1 | 0.00 | phoneme | stimuli/audio/lw1_0.wav | t_B | NaN | lw1 | sentence | 205.0 | Allison |
| 2 | 0.00 | word | stimuli/audio/lw1_0.wav | NaN | Tara | lw1 | sentence | 205.0 | Allison |
| 3 | 0.08 | phoneme | stimuli/audio/lw1_0.wav | eh_I | NaN | lw1 | sentence | 205.0 | Allison |
| 4 | 0.17 | phoneme | stimuli/audio/lw1_0.wav | r_I | NaN | lw1 | sentence | 205.0 | Allison |
| ... | ... | ... | ... | ... | ... | ... | ... | ... | ... |
| 3129 | 51.85 | phoneme | stimuli/audio/lw1_3.wav | p_I | NaN | lw1 | sentence | 205.0 | Allison |
| 3130 | 51.94 | phoneme | stimuli/audio/lw1_3.wav | iy_I | NaN | lw1 | sentence | 205.0 | Allison |
| 3131 | 52.03 | phoneme | stimuli/audio/lw1_3.wav | sh_I | NaN | lw1 | sentence | 205.0 | Allison |
| 3132 | 52.11 | phoneme | stimuli/audio/lw1_3.wav | iy_I | NaN | lw1 | sentence | 205.0 | Allison |
| 3133 | 52.12 | phoneme | stimuli/audio/lw1_3.wav | z_E | NaN | lw1 | sentence | 205.0 | Allison |

**Figure 4.** MEG data annotations: Pandas DataFrame of sound, phoneme and word time-stamps.